\documentclass[twocolumn]{aastex631}

\usepackage{amsmath}

\graphicspath{{./}{figures/}}

\begin{document}
\title{Towards Measuring Microlensing Event Rate in the Galactic Center: I. Events Detection from the UKIRT Microlensing Survey Data}

\correspondingauthor{Bo Ma}
\email{mabo8@mail.sysu.edu.cn}

\author[0000-0002-0786-7307]{Yongxin Wen}
\affiliation{School of Physics and Astronomy, \\ Sun Yat-sen University, Zhuhai 519082, China; {\it mabo8@mail.sysu.edu.cn}\\
}
\affiliation{CSST Science Center for the Guangdong-HongKong-Macau Great Bay Area, \\ Sun Yat-sen University, Zhuhai 519082, China;\\}

\author{Weicheng Zang}
\affiliation{Center for Astrophysics | Harvard \& Smithsonian 60 Garden St., Cambridge, MA 02138, USA}

\author[0000-0002-0378-2023]{Bo Ma}
\affiliation{School of Physics and Astronomy, \\ Sun Yat-sen University, Zhuhai 519082, China; {\it mabo8@mail.sysu.edu.cn}\\
}
\affiliation{CSST Science Center for the Guangdong-HongKong-Macau Great Bay Area, \\ Sun Yat-sen University, Zhuhai 519082, China;\\}

\begin{abstract}
To overcome the high optical extinction, near-infrared observations are needed for probing the microlensing events toward the Galactic center. The 2015-2019 UKIRT microlensing survey toward the Galactic center is the first dedicated precursor near-infrared (NIR) survey for the Nancy Grace Roman Space Telescope. We here analyze the online data from the UKIRT microlensing survey, reaching $l=b=0^\circ$. Using the event-finder algorithm of KMTNet with the $\Delta \chi^2$ threshold of 250, we find 522 clear events, 436 possible events, and 27 possible anomalous events. We fit a point-source point-lens (PSPL) model to all the clear events and derive the PSPL parameters with uncertainties using a Markov chain Monte Carlo method. Assuming perfect detection efficiency, we compute the uncorrected event rates, which should serve as the lower limits on the true event rate. We find that the uncorrected NIR event rates are likely rising toward the Galactic center and higher than the optical event rates. 
\end{abstract}


\section{Introduction} \label{sec:intro}
Since the first exoplanet was detected via the gravitational microlensing technique \citep{Mao91, Andy1992, Bennett1996} two decades ago \citep{Bond04}, microlensing has established itself as an important technique for bound planets near and beyond the snow line \citep{Min2011} and free-floating planets (FFPs). For bound planets, the statistical samples from follow-up observations \citep{Gould10, Cassan2012} and wide-field high-cadence surveys \citep{Wise, Suzuki2016}, which contains ${\cal O}(10^1)$ planets, have shown the abundance of bound planets near and beyond the snow line, with an occurrence frequency of $\sim 50\%$ to 100\% (with planetary masses down to several Earth-mass). The new-generation microlensing survey, the Korea Microlensing Telescope Network (KMTNet, \citealt{KMT2016}), began its regular observations in 2016. The KMTNet is expected to form a statistical sample including ${\cal O}(10^2)$ bound planets with the newly developed AnomalyFinder algorithm \citep{OB191053, 2019_prime} and first yield the planetary occurrence frequency at about one Earth-mass. For FFPs, the fourth phase of the Optical Gravitational Lensing Experiment (OGLE, \citealt{Udalski1994, OGLEIV}) and KMTNet have detected ${\cal O}(10^1)$ candidates (\citealt{Mroz2017a, KB172820} and references therein). If these candidates are real FFPs, low-mass FFPs (from Earth-mass to Neptune-mass) are more common than stars \citep{Mroz2017a, Gould2022}. 

Thanks to the continuous and stable observations from space, the upcoming space missions, the Nancy Grace Roman Space Telescope \citep[{\it Roman}, former {\it WFIRST}, ][]{Spergel2015} will detect ${\cal O}(10^3)$ bound and free-floating planets, with planetary masses down to Mars-mass \citep{Penny19, Johnson2020}. Moreover, {\it Roman} 's high-resolution imaging can yield the mass function for bound planets, and the satellite microlensing parallax \citep{1966MNRAS.134..315R, 1994ApJ...421L..75G, Gould1995single, Yan22, Bachelet22b} from {\it Roman} and other space telescope can also constrain the mass function for bound and free-floating planets. Up to now, the ground-based observations have only yielded the planet/host mass-ratio ($q$) function for bound planets, because generally the light-curve analysis only well determines $q$ but the mass of the host remains unknown. Moreover, the only method to determine the mass of FFPs is observing FFPs from two wide-field telescopes separated by $D\sim{\cal O}(0.01)$AU \citep{CMST}. 

The yield of the {\it Roman} microlensing survey depends heavily on the microlensing event rate per star per year, $\Gamma$, for the targeted fields. \citet{Sumi13} and \citet{Sumi16} found that the event rate is increasing toward low Galactic latitudes using 474 microlensing events at $b = -6^\circ \sim -1.5^\circ$ detected by the Microlensing Observations in Astrophysics (MOA, \citealt{Sako2008}) survey. Using a homogeneous sample of $\sim 8000$ events found by the OGLE-IV survey, \cite{Mroz19} confirmed this trend and found that the event rates turn over at $b \sim -1.5^{\circ}$, which is consistent with the prediction based on the Besançon model of the Galaxy \citep{Besancon_model, Awiphan2016}. This turnover is caused by high interstellar extinction at low Galactic latitudes. The MOA and OGLE surveys use the optical bands, in which the number of observable sources and potential lenses sharply decreases in these regions because observable sources are located closer. In the near-infrared (NIR) band, which is the band of {\it Roman}, the turnover may not exist due to more and farther observable sources. For example, using 360 microlensing events detected by the VISTA Variables in the V$\acute{\rm i}$a L$\acute{\rm a}$ctea (VVV) survey \citep{VVV}, \cite{Navarro20} found that the number of observable stars and the microlensing events increases rapidly toward the Galactic center.

To date, the microlensing event rate in the NIR band is still blank. The current predictions of the {\it Roman} microlensing survey were based on the event rates in the optical bands and thus the predicted yield at $|b| < 1^{\circ}$ is low \citep{Penny19}. 
To provide microlensing event rates in the NIR band in the Galactic center, we initiate a project to analyze the ground-based NIR surveys. As the first step (i.e., Paper I), in this work we search for microlensing events from the data taken by the United Kingdom Infrared Telescope (UKIRT) microlensing survey \citep{Shvartzvald17, Shvartzvald18}, and calculate the raw event rate by assuming a perfect detection efficiency \citep{Shvartzvald17}. The microlensing event selection criteria adopted by our algorithm in this paper will be applied in a follow-up study (i.e., Paper II), in which we will calculate the event detection efficiency of the UKIRT microlensing survey using simulated survey data and estimate the true NIR event rate towards Galactic center.
During the year 2015--2019, UKIRT conducted a dedicated NIR microlensing survey toward the Galactic bulge, and the utility of the UKIRT NIR microlensing survey has been demonstrated by the analysis of part of the UKIRT microlensing data, including five new microlensing events from the 2016 data \citep{Shvartzvald17}, a giant planet solely detected by UKIRT at $(\ell, b) = (-0.12, -0.33)$ \citep{Shvartzvald18}, and the source color measurements for some planetary events (e.g., \citealt{OB161190, OB170173}). 

The paper is structured as follows. We introduce the observations and the online data of the UKIRT microlensing survey in Section~\ref{sec:Observations}. In Section~\ref{sec:method}, we present the microlensing events detection algorithm and the results. Finally, we discuss the implications of our results in Section~\ref{sec:discussion} and provide a conclusion in Section~\ref{sec:conclusion}.

\section{UKIRT Microlensing Survey Data} \label{sec:Observations}

The UKIRT microlensing survey data were taken using the Wide Field Camera (WFCAM) with a pixel scale of $0.4''$ and a field of view (FoV) of about 0.8 square degrees \citep{Shvartzvald17, Shvartzvald18}. We refer the reader to \citet{Casali07} for more details about the instrument and telescope. 

\begin{figure*}[ht!]
\plotone{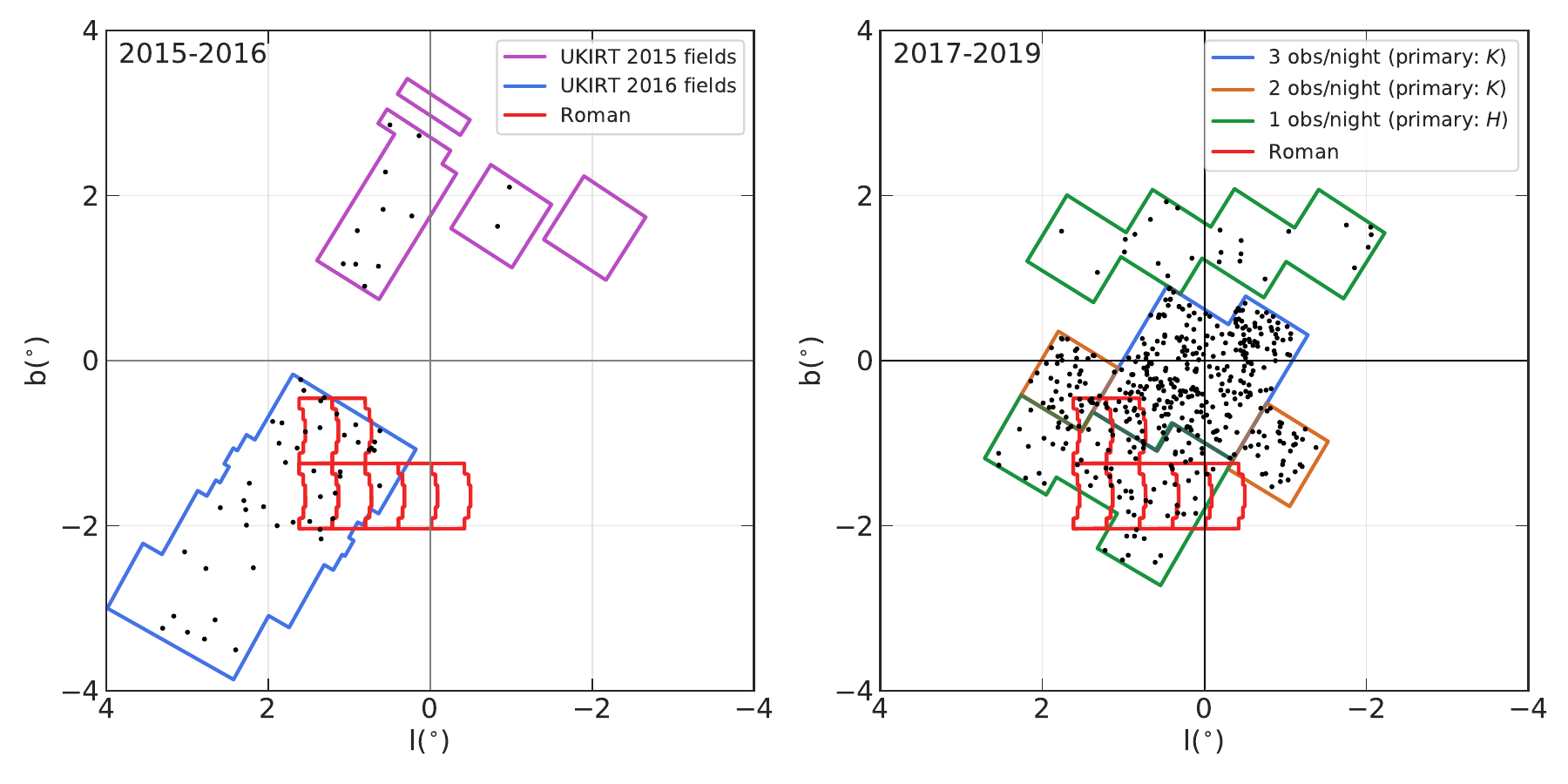}
\caption{UKIRT microlensing survey fields from 2015 to 2019 in Galactic coordinate. The UKIRT 2015 (purple) and 2016 (blue) microlensing fields are shown in the left panel, and UKIRT 2017-2019 survey fields are shown in the right panel. In the right panel, the green fields with positive Galactic latitude are designated as northern Galactic bulge fields, blue fields are designated as the central Galactic bulge fields, and both orange fields and the green fields with negative Galactic latitude are designated as southern Galactic bulge fields. Please see the UKIRT Coverage Maps and Magnitude Ranges page (\url{https://exoplanetarchive.ipac.caltech.edu/docs/UKIRT_figures.html}) for more details. The black dots mark the positions of clear microlensing events found by this work. The red curves represent the
footprints of the seven provisional baseline {\it Roman} Cycle 7 fields \citep{Penny19}. 
\label{fig:ukirt_fields}}
\end{figure*}

\begin{figure}[ht!]
\epsscale{1.15}
\plotone{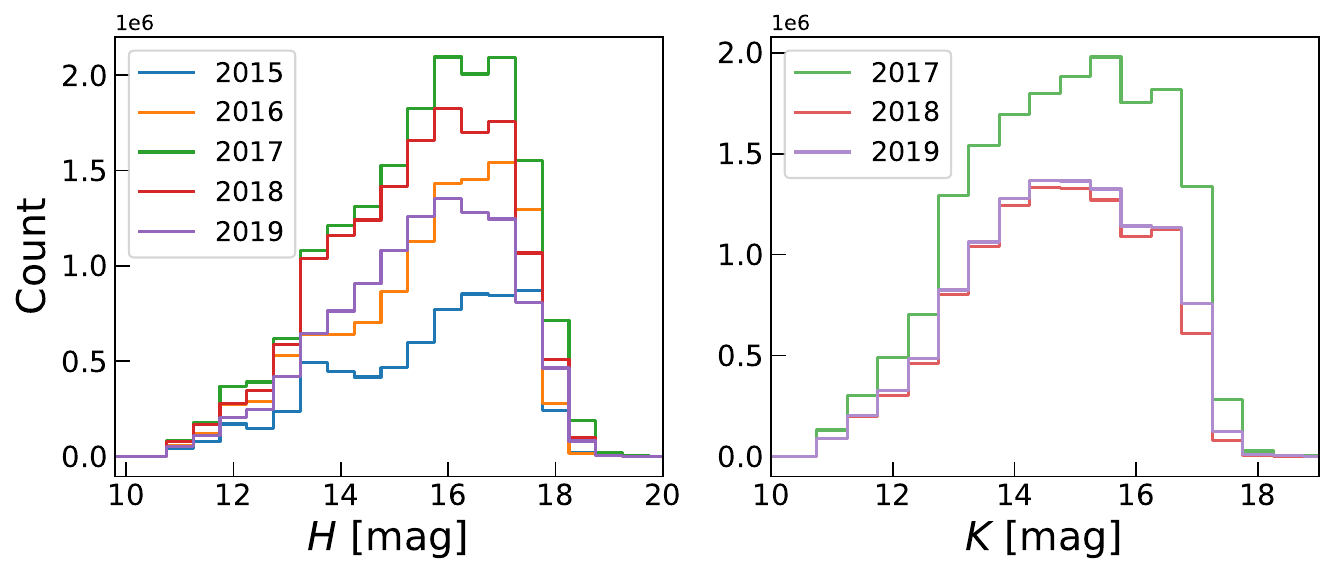}
\caption{Magnitude distribution of UKIRT microlensing survey targets in $H$-band and $K$-band.
\label{fig:magnitudes}}
\end{figure}

\begin{figure}[ht!]
\epsscale{1.2}
\plotone{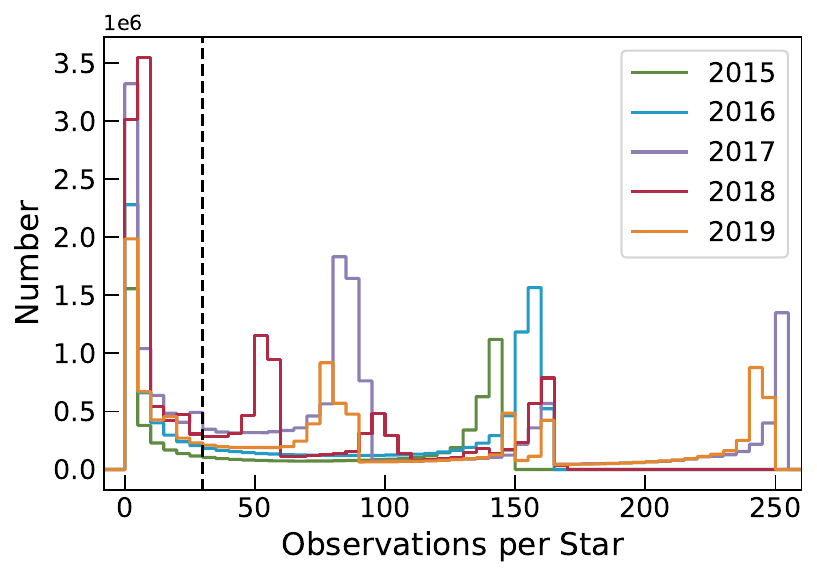}
\caption{Distribution of number of observation epochs per target in the 2015-2019 UKIRT microlensing survey. The black dashed line corresponds to 30 epochs per target, which is the threshold cut required for the microlensing event search in this work.
\label{fig:obs_per_star}}
\end{figure}

The UKIRT microlensing survey fields from 2015-2019 are summarized in Figure~\ref{fig:ukirt_fields}. In 2015, UKIRT started a microlensing survey of the inner Galactic bulge region to support the {\it Spitzer} microlensing project \citep{YeeSpitzer}, which lasted 39 nights with a nominal cadence of 5 epochs per night in $H$ band \citep{Shvartzvald17}. This was subsequently followed by the 2016 campaign, which provided a simultaneous NIR coverage of the K2 Campaign 9 survey \citep{Henderson16} area over 91 nights \citep{Shvartzvald17}. 2017 marked the beginning of the dedicated {\it Roman} precursor survey using UKIRT. The 2017, 2018 and 2019 UKIRT campaign obtained 131, 116 and 146 nights observations over 10.5~$\rm deg^2$ field of view in both $H$ band and $K$ band.

The UKIRT survey data were reduced using the CASU multi-aperture photometry pipeline, which can produce calibrated 2MASS $H$-band and $K$-band magnitudes in very crowded stellar fields \citep{2004SPIE.5493..411I, 2009MNRAS.394..675H}.
According to \citet{Shvartzvald18}, two methods have been applied here in reducing UKIRT microlensing survey data, (a) Two Micron All Sky Survey (2MASS) calibrated soft-edge aperture photometry, and (b) 2MASS-calibrated point-spread function photometry using SExtractor \citep{Bertin96} and PSFEx \citep{Bertin11}. \citet{Hajdu20} have found issues with the absolute photometric calibration when reducing VVV survey data using CASU pipeline and the 2MASS catalog. However, this will not affect our study here since detecting microlensing events from light curves does not rely on an absolute magnitude scale. \citet{Hajdu20} have also found large scatter caused by the CASU pipeline when dealing with crowded field. 
To evaluate the the quality of the UKIRT photometry, we have conducted an analysis of the percentage of UKIRT light curves having larger than usual scatter. 
We first calculate the standard deviation (STD) of all photometric data points for each light curve.
Then we divide all light curves into magnitude bins, each with an interval of 0.5~magnitudes. Within these bins, we calculate the median STD value, denoted as STD$_{\rm median}$, for all light curves in each bin. The percentage of light curves exhibiting STD values larger than 5 times STD$_{\rm median}$ within each magnitude bin is presented in Table~\ref{tab:bad_lightcurve}. 
We find that generally fewer than 5$\%$ of the light curves are suffering larger than usual scattering within each magnitude bin. When accounting for the different number of light curves in each magnitude bin, the overall percentage is 2.5$\%$ and 2.2$\%$ in K- and H-band respectively for the entire dataset, which will not seriously impact the detection of microlensing events from these light curves.

\begin{deluxetable}{cccc}
\tablewidth{0pt}
\caption{Percentage of the UKIRT light curves having standard deviation 5 times larger than the median standard deviation in each magnitude bin. 
We only show the statistics for targets from 13 to 18~mag, which dominate the whole UKIRT microlensing sample. \label{tab:bad_lightcurve}
}
\tablehead{
\colhead{mag} & \colhead{H-band} & \colhead{K-band} 
}
\startdata
{} 13.0-13.5 & 6.3$\%$ & 4.9$\%$  \\
{} 13.5-14.0 & 5.1$\%$ & 3.9$\%$  \\
{} 14.0-14.5 & 4.0$\%$ & 3.1$\%$  \\
{} 14.5-15.0 & 3.3$\%$ & 2.5$\%$  \\
{} 15.0-15.5 & 2.6$\%$ & 1.8$\%$  \\
{} 15.5-16.0 & 2.1$\%$ & 1.2$\%$  \\
{} 16.0-16.5 & 1.6$\%$ & 0.6$\%$  \\
{} 16.5-17.0 & 0.9$\%$ & 0.4$\%$  \\
{} 17.0-17.5 & 0.5$\%$ & 0.4$\%$  \\
{} 17.5-18.0 & 0.2$\%$ & 0.5$\%$  \\
\enddata
\end{deluxetable}

In total, the five-year-long survey produced 100.5~million light curves for $\sim66$~million targets \citep{ukirt_data}, with each target yielding one light curve during annual observations. 
We download all the light curve products from the UKIRT Microlensing Survey website\footnote{\url{https://exoplanetarchive.ipac.caltech.edu/docs/UKIRTMission.html}}. The $H$- and $K$- band magnitude distributions for all the survey targets are shown in Figure~\ref{fig:magnitudes}. 
The majority of the survey targets have magnitudes brighter than 18 in both the $H$ and $K$ band, with the photometric precision of 0.01-0.1~mag. The distribution of observing epochs for each year is illustrated in Figure~\ref{fig:obs_per_star}. The number of accumulated epochs varies each year due to the different duration of each observing season.

\section{Methods and Results} \label{sec:method}

\subsection{Detection of Microlensing Events}
\label{sec:detection_algorithm}

\begin{figure}[ht!]
\epsscale{1.2}
\plotone{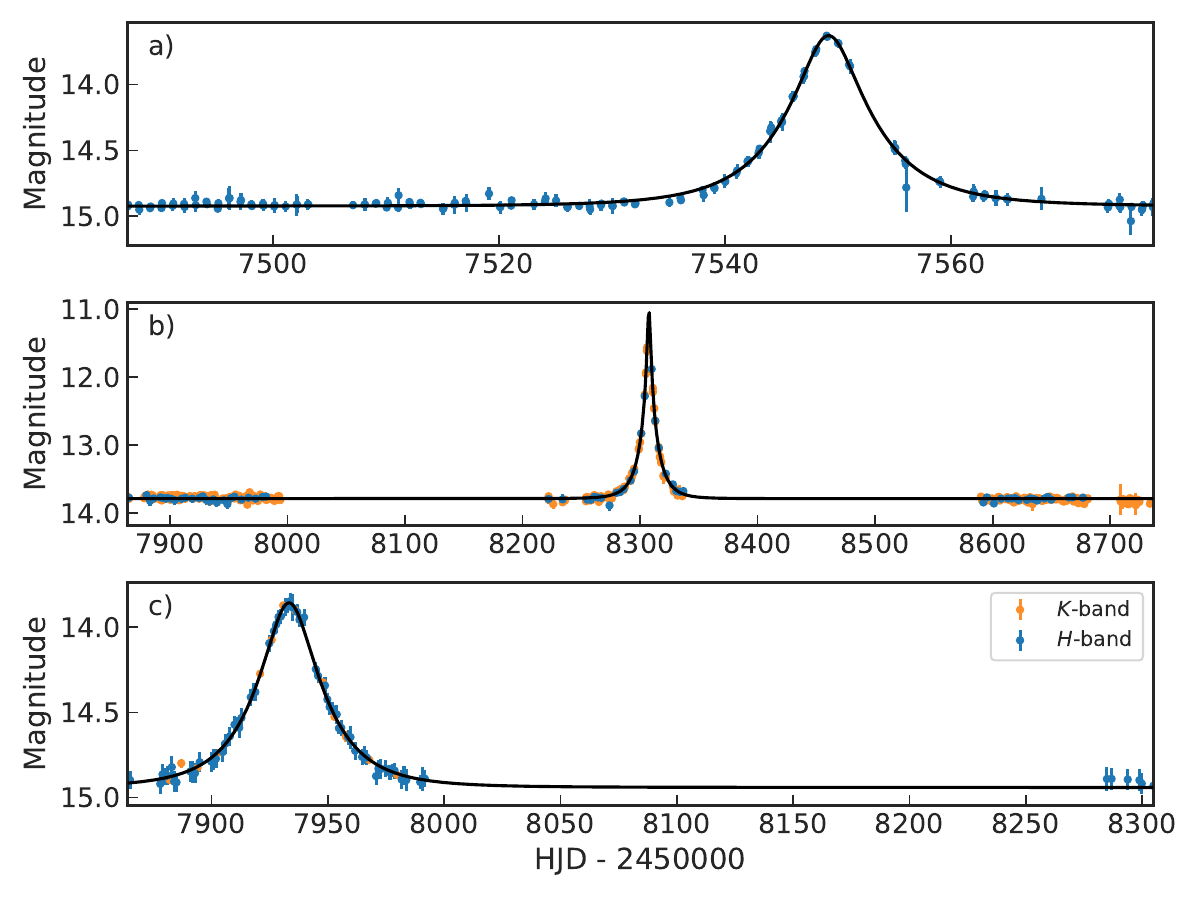}
\caption{Three examples of clear UKIRT microlensing events found by our search. Top panel: a microlensing event found in the 2016 UKIRT survey data. Middle panel: a microlensing event dounf in the 2018 UKIRT survey data. The 3-year combined data show a clear baseline.  Bottom panel: a microlensing event found in the northern Galactic bulge fields in 2017. The 2018 data points are crucial as they present a clear baseline. If the 2018 data were missing, we would classify this event as a possible event. 
\label{fig:examples}}
\end{figure}

Because the UKIRT microlensing survey targeted different Galactic bulge regions each year, we decide to perform an independent event search for each year's data. The light curves with less than 30 data points are excluded from this analysis. We employ the ``completed-event'' microlensing event-finder algorithm of \cite{Kim_2018} to identify possible microlensing events in these light curves, which adopts a pre-calculated \citet{Gould96} two-dimensional grid of $(t_0, t_{\rm eff})$ to locate potential microlensing events in the light curves, 
\begin{equation} \label{eqn:event_finder}
\begin{split}
F(t) & = f_1 A_j[Q(t;t_0,t_{\rm eff})] + f_0; \\
 Q(t;t_0,t_{\rm eff}) & \equiv 1 + \biggl(\frac{t-t_0}{t_{\rm eff}}\biggr)^2; \\
(j & =1,2) \\
\end{split}
\end{equation}
where
\begin{equation} \label{eqn:ajoftau}
\begin{split}
A_{j=1}(Q) & = Q^{-1/2}; \\
A_{j=2}(Q) & = \frac{Q + 2}{\sqrt{Q(Q+4)}} \\
           & = [1 - (Q/2 +1)^{-2}]^{-1/2}, \\
\end{split}
\end{equation}
where $t_{\rm eff}$ is the effective timescale, $t_0$ is the time of maximum magnification, $F(t)$ is the observed flux, and $(f_1, f_0)$ are the two flux parameters, which are derived by a linear fit. We restrict the fits to photometric data within $t_0\pm Zt_{\rm eff}$, where $Z=5$ and calculate $\chi_{\rm \mu lens}$ from this fit. The significance of the potential microlensing signal is evaluated using 
\begin{equation}
\Delta \chi^2 = (\frac{\chi_{\rm flat}^2}{\chi_{\rm \mu lens}^2}-1)N_{\rm data},
\end{equation}
where $\chi_{\rm flat}$ is calculated using a flat light-curve model fitting to the data. As the UKIRT microlensing survey is not a high-cadence survey like KMTNet, we set the $\Delta \chi^2$ threshold as 250. 

For each candidate event, we perform a static point-source point-lens (PSPL, \citealt{Paczynski86}) model fitting. Because the 2017-2019 UKIRT survey was conducted toward the same sky area, we decide to apply the PSPL model fitting to the 3-yr combined light curve data of the same star, which has a time span of $\sim$900~days and two 200$-$day gaps in between. A static PSPL model has three parameters, $t_0$, $u_0$, and $t_{\rm E}$, which respectively represent the time of lens-source closest approach, the minimum impact parameter scaled to the Einstein radius, and the Einstein radius crossing time. Then, the lens-source projected separation normalized by the Einstein radius, $u(t)$, can be described by 
\begin{equation} \label{eq:4}
u(t) = \sqrt{u_0^2+\frac{(t-t_0)^2}{t_{\rm E}^2}},
\end{equation}
and the the magnification, $A(t)$, can be computed by 
\begin{equation} \label{eq:3}
A(t) = \frac{u(t)^2+2}{u(t)\sqrt{u(t)^2+4}}.
\end{equation}
The observed flux $F(t)$, is modeled as 
\begin{equation}
    F(t) = F_{\rm S} A(t) + F_{\rm B},
\end{equation}
where $F_{\rm S}$ and $F_{\rm B}$ represent the source flux and any blended flux, respectively. We introduce the blending parameter, $f_{bl}$, for the strength of the blending, and 
\begin{equation} \label{eq:2}
f_{bl} = \frac{F_{\rm S}}{F_{\rm S}+ F_{\rm B}}.
\end{equation}
Here $f_{bl} = 1$ means no blending and $ f_{bl} \sim 0$ means extreme blending. 

We use the MulensModel code \citep{Poleski_2019} and the Nelder-Mead algorithm to find the best-fit parameters and then employ the Markov chain Monte Carlo (MCMC) technique by utilizing the emcee sampler code from \citet{2013PASP..125..306F} to estimate the uncertainties of the model parameters. 

The final step is a vetting test of all the candidate events, including checking for a good fit to the standard PSPL model, a clear constant baseline, and at least one data point in the rising and falling part of the light curve. We separate all the selected microlensing events into two different categories according to our vetting results: a sample of clear detection events and a sample of possible detection events. For clear events, they meet all of the vetting criteria above. For possible events, they do not meet all of the criteria but are still likely to be real microlensing events. We also exclude duplicated events whose angular distances are less than 1 arcsec apart. As a result, we find a total of 522 clear events and 436 possible events. We display an example clear event light curve with a fitted PSPL model in panel (a) of Figure~\ref{fig:examples}. We note that the five new microlensing events found by \citet{Shvartzvald17} using the 2016 UKIRT data are all identified as clear events in this work.

\begin{figure*}[ht!]
\plotone{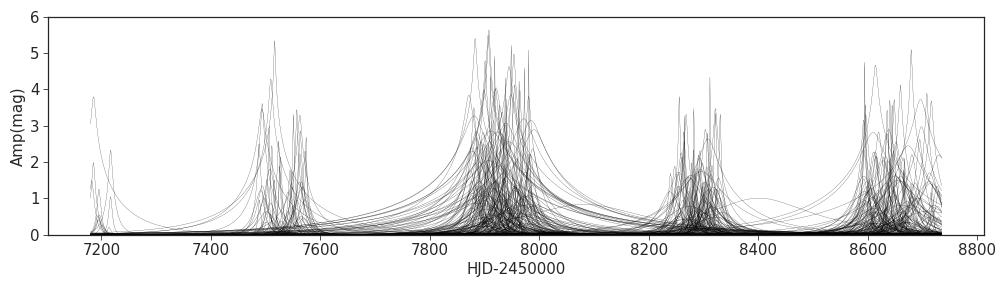}
\caption{The best-fit PSPL model curves for all UKIRT clear microlensing events detected between 2015 and 2019. 
\label{fig:amp_mag_2017-2019}}
\end{figure*}

\begin{figure}[ht!]
\epsscale{1.2}
\plotone{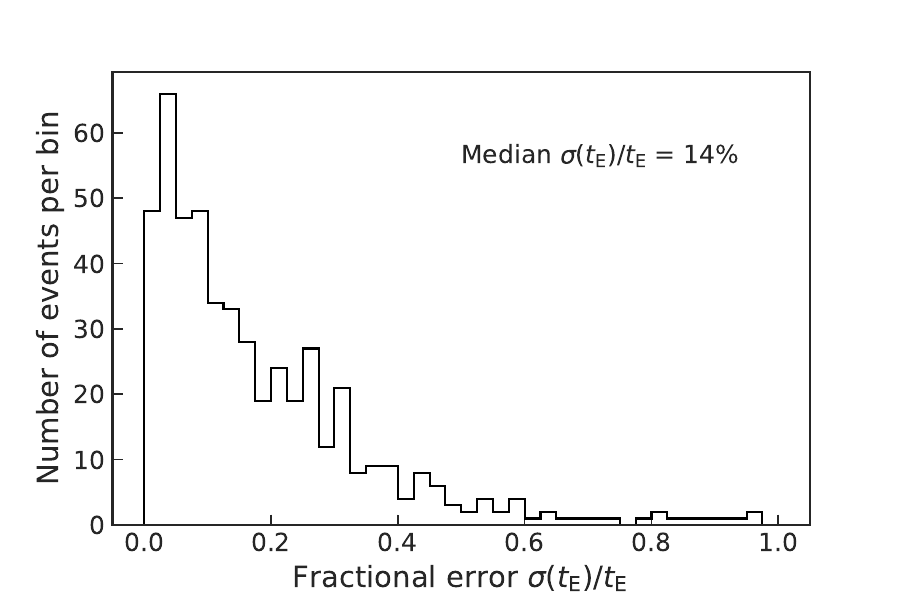}
\caption{Distribution of fractional uncertainties of Einstein timescales of clear microlensing events from the UKIRT survey. The median uncertainty is 14$\%$, while $\sigma(t_{\rm E})/t_{\rm E} \le 0.5$ for 91$\%$ of events in the sample.
\label{fig:sigma_tE_tE}}
\end{figure}

The number of events detected each year are summarized in Table~\ref{tab:events_in_each_year}, and the spatial distribution of all detected clear events is shown in Figure~\ref{fig:ukirt_fields}. The parameters of all the detected events, including the PSPL model fitting results and corresponding $1\sigma$ uncertainties, are summarized in Table~\ref{tab:event_paramenter}. The fitted PSPL model light curves for all the clear events are shown in Figure~\ref{fig:amp_mag_2017-2019}. 
All the events have a median Einstein timescale of $t_{\rm E}$ of $24.4$~days.
Figure~\ref{fig:sigma_tE_tE} shows the distribution of fractional uncertainties of Einstein timescales. The median uncertainty is 14$\%$, and $\sigma(t_{\rm E})/t_{\rm E} \le 0.5$ for 91$\%$ of events in the clear event sample.

It is obvious that a longer observation time span, higher cadence, and larger survey fields allow the detection of more microlensing events by comparing results from the 2015-2016 campaign with the 2017-2019 campaign. The longer observation time span for 2017--2019 not only allows a clearer baseline for the fit of short-timescale events but also allows the detection of long-timescale events, which is evident in the panel (b) and (c) of Figure~\ref{fig:examples}. 
Due to irregular sampling in the UKIRT survey, there are gaps in each year's data, with 2018 having a relatively higher number of gaps. As a result, despite an observation time span of 116 days, the number of events discovered in 2018 was significantly lower than that in the other two years. Notably, the number of events detected in 2017 was the highest, owing to the inclusion of a substantial amount of data from the northern bulge. 

\begin{deluxetable}{cccc}
\tablewidth{0pt}
\caption{Summary of the UKIRT microlensing events in Different Years. 
\label{tab:events_in_each_year}}
\tablehead{
\colhead{Year} & \colhead{Observing nights} & \colhead{ Clear events} & \colhead{Possible events}
}
\startdata
{} 2015 & 39 & 12 & 13 \\
{} 2016 & 91 & 47 & 51 \\
{} 2017 & 131 & 202 & 205 \\
{} 2018 & 116 & 102 & 43 \\
{} 2019 & 146 & 159 & 124 \\
\enddata
\end{deluxetable}

\subsection{Possible Anomalous Microlensing Events}

\begin{figure*}[ht!]
\includegraphics[width=1\textwidth]{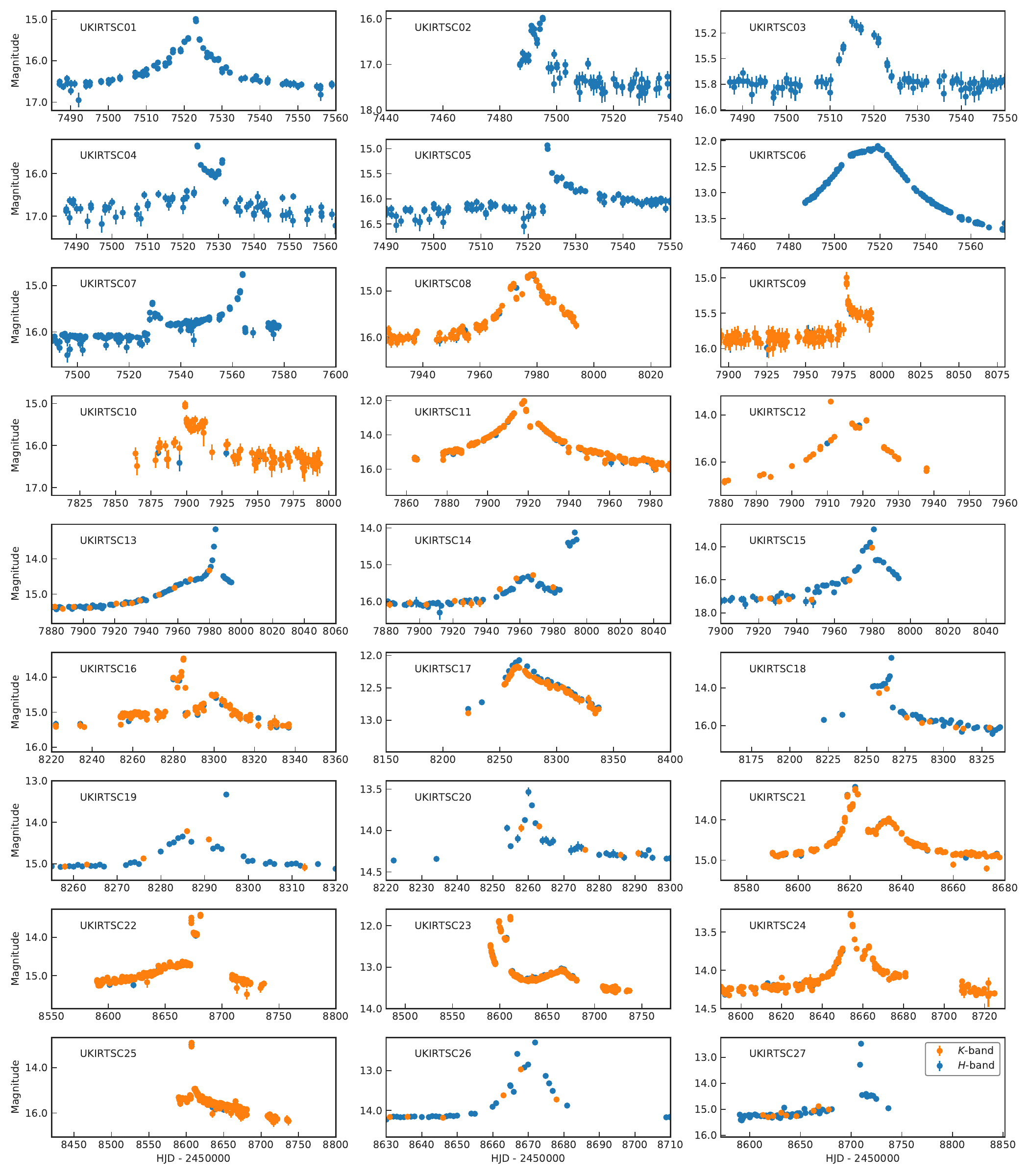}
\caption{Possible anomalous microlensing events detected from the 2015-2019 UKIRT microlensing survey. The $H$- and $K$- band data are marked using blue and orange dots, respectively. Events UKIRTSC01 to UKIRTSC07 are detected in the survey data from 2015 to 2016, whereas the remaining events are detected in the survey data from 2017 to 2019. The ID of each event is indicated in the upper left corner of the corresponding panel. The time axis of each panel is selected so that light curve data near $t_0$ are displayed right in the middle of the plot. Although most of the events have been monitored for three years, only light curve data near the time of occurrence of each event are displayed.
\label{fig:special1}}
\end{figure*}

We also identified 27 possible anomalous microlensing events, which are summarized in Table~\ref{tab:special_cases} and Figure~\ref{fig:special1}. These events do not have a good fit for the standard PSPL model, so we do not consider them as clear events. These events are probably binary microlensing events, i.e., binary-lens single-source (2L1S) or single-lens binary-source (1L2S) events. If the 27 events are all real anomalous microlensing events, the binary ratio is $\sim 5\%$, which is lower than the $\sim 11\%$ binary ratio of the KMTNet survey with the AnomalyFinder algorithm \citep{2019_prime,2018_prime,2018_subprime,2019_subprime,2016_prime}. This is reasonable because the KMTNet survey has a much higher cadence than the UKIRT survey. See Figure 12 of \cite{Kim_2018} for the KMTNet field placement. The analysis of these events is beyond the scope of this work, and we encourage colleagues in the community to investigate these events in the future. 

Previously, \citet{Shvartzvald18} detected a giant planet UKIRT-2017-BLG-001Lb using the 2017 UKIRT survey data, with a planet-to-host mass ratio of $q=1.50^{+0.14}_{-0.17}\times10^{-3}$. This planet cannot be detected by an optical survey because the event lies at $0.35^{\circ}$ from the Galactic center and suffers from an extremely high extinction of $A_K = 1.68$. We have also identified this anomalous event using our algorithm and labeled it as UKIRTSC11. Its light curve is also shown in Figure~\ref{fig:special1}.

\begin{deluxetable*}{cccccccc}
\tablewidth{0pt}
\caption{Information of the possible anomalous microlensing events from the UKIRT Survey. \label{tab:special_cases}}
\tablehead{
\colhead{ID} & \colhead{RA} & \colhead{DEC}  & \colhead{KMTNet ID} & \colhead{MOA ID} & \colhead{OGLE ID}  \\
\colhead{} & \colhead{(deg)} & \colhead{(deg)}  & \colhead{} & \colhead{} & \colhead{}
}
\startdata
UKIRTSC01&268.604794&-27.700445&-&-&-\\
UKIRTSC02&271.392181&-27.660869&KB160042&-&OB160562\\
UKIRTSC03&269.572005&-28.681166&-&-&-\\
UKIRTSC04&269.641109&-27.378776&KB160617&-&OB160887\\
UKIRTSC05&268.710515&-28.318638&KB160180&-&-\\
UKIRTSC06&267.586949&-29.315448&KB160592&-&OB160262\\
UKIRTSC07&268.297288&-27.709069&-&-&-\\
UKIRTSC08&265.691349&-29.220871&-&-&-\\
UKIRTSC09&266.821881&-29.645386&-&-&-\\
UKIRTSC10&267.293634&-29.655658&-&-&-\\
UKIRTSC11&266.654085&-29.211372&-&-&-\\
UKIRTSC12&267.285966&-27.952637&-&-&-\\
UKIRTSC13&268.536294&-27.754706&-&-&-\\
UKIRTSC14&268.745669&-28.439389&KB170213&MB17431&OB171323\\
UKIRTSC15&267.850449&-29.291020&KB170318&MB17423&OB171471\\
UKIRTSC16&266.336611&-28.212144&-&-&-\\
UKIRTSC17&267.742759&-27.816703&-&-&-\\
UKIRTSC18&268.118472&-28.393161&-&-&OB180752\\
UKIRTSC19&268.157462&-28.603910&KB182095&-&OB181055\\
UKIRTSC20&267.944578&-28.211693&-&-&OB180856\\
UKIRTSC21&265.455809&-29.671021&-&-&-\\
UKIRTSC22&265.651981&-29.506094&-&-&-\\
UKIRTSC23&265.662585&-28.994653&-&-&-\\
UKIRTSC24&265.366570&-29.000787&-&-&-\\
UKIRTSC25&267.537198&-28.331357&-&-&-\\
UKIRTSC26&268.935118&-29.752919&-&MB19277&OB190950\\
UKIRTSC27&268.663956&-29.104072&KB191450&MB19363&OB191048\\
\enddata
\tablecomments{Events names of OGLE, MOA and KMTNet are abbreviations, e.g., KMT-2016-BLG-0042 to KB160042.}
\end{deluxetable*}

\subsection{Raw NIR Event Rate}

The microlensing event rate in the NIR band is the key scientific goal of the UKIRT microlensing survey. Detailed simulations of the detection efficiency for the UKIRT data are needed to obtain the underlying true event rate, which is beyond the scope of this paper. However, we still can estimate the distribution of raw event rates by a catalog-level calculation. Assuming a perfect detection efficiency, \citet{Shvartzvald17} calculated the raw event rate using the number of events identified, the total number of light curves obtained, and the survey duration by the following expression 
\begin{equation}
\Gamma_{\rm raw} = \frac{N_{\rm events}}{N_{\rm tot}T_{\rm obs}}, \label{eq:rate}
\end{equation}
where $N_{\rm tot}$ is the number of light curves searched, $N_{\rm events}$ is the total event number, and $T_{\rm obs}$ is the time span of the survey. We note that $\Gamma_{\rm raw}$ should be the lower limits on the true event rates due to the inefficiency of detecting all events. Then, by analyzing eight events from about three million light curves, \citet{Shvartzvald17} found that the raw NIR event rate is $96_{-33}^{+47}\times10^{-6}$~star$^{-1}$~yr$^{-1}$ near $l = 1^\circ, b = -0.75^\circ$. 

\begin{figure*}[ht!]
\centering
\includegraphics[width=0.9\textwidth]{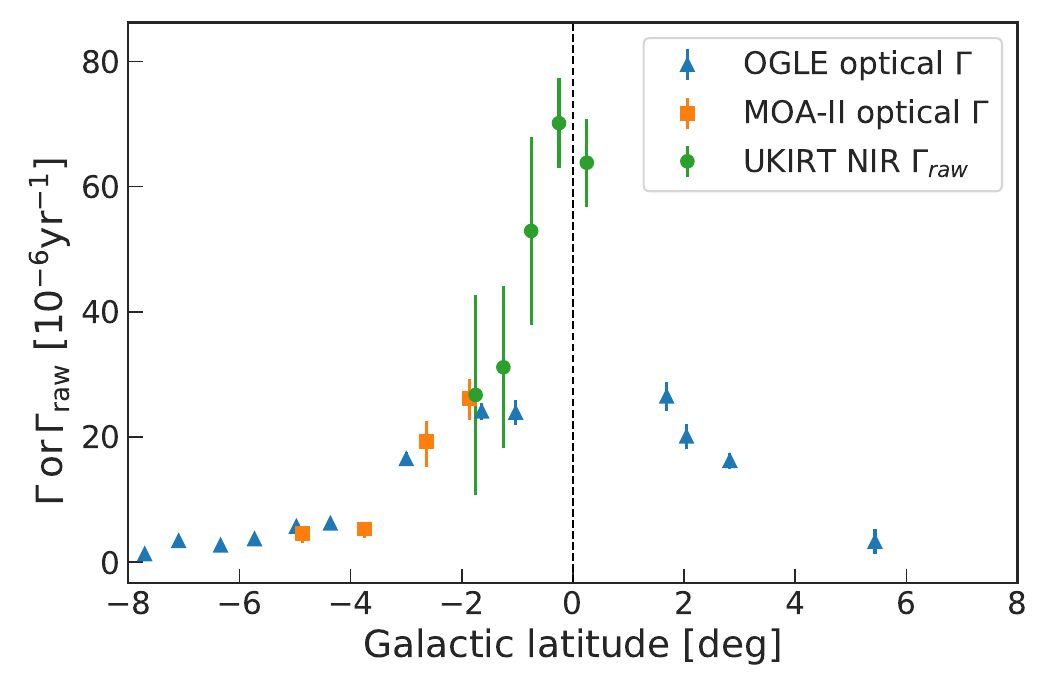}
\caption{The raw NIR event rate distribution along the galactic latitude derived from 2017--2019 UKIRT clear events. The MOA-II optical event rate are drawn using data from Table~4 of \citet{Sumi16}, and the OGLE optical event rate are drawn using data from Table~7 of \citet{Mroz19}. 
All the event rate data presented is limited to regions where $|l|<1^\circ$.
The error bars of the UKIRT raw event rate are calculated using Poisson counting statistics. The UKIRT raw event rates are also presented in Table \ref{tab:raw_event_rate}.
\label{fig:event_rate}}
\end{figure*}

Following the method of \citet{Shvartzvald17} and using all clear UKIRT events from 2017--2019, we calculate the raw event rates and display the distribution along the Galactic latitude in Figure~\ref{fig:event_rate}, in which we limit the calculation to regions at $|l|<1^\circ$ to aid the comparison with previous works. The raw event rates are summarized in Table~\ref{tab:raw_event_rate}.

The true event rate distributions from the OGLE survey \citep{Mroz19} and the MOA-II survey \citep{Sumi16} are also shown in Figure~\ref{fig:event_rate}, and the raw UKIRT event rates are in agreement with the OGLE and MOA-II event rates within $1\sigma$ around $b = -1^\circ$. In addition, there is probably a rising trend in the NIR event rates toward the low Galactic latitude area.

\begin{deluxetable}{cccc}
\tablewidth{0pt}
\caption{Raw Microlensing Event Rates from the UKIRT Microlensing Survey. \label{tab:raw_event_rate} }
\tablehead{
Region Number & \colhead{l} & \colhead{b} & \colhead{$\Gamma_\mathrm{raw}$} \\
& \colhead{(deg)} & \colhead{(deg)} & \colhead{$\rm (10^{-6}~yr^{-1})$}
}
\startdata
1 & \lbrack -1, 1] &\lbrack-2.0, -1.5] & $26.7\pm16.0$ \\
2 & \lbrack -1, 1] &\lbrack-1.5, -1.0] & $31.1\pm13.0$ \\
3 & \lbrack -1, 1] &\lbrack-1.0, -0.5] & $52.9\pm15.0$ \\
4 & \lbrack -1, 1] &\lbrack-0.5, \;0.0] & $70.1\pm\;7.2$  \\
5 & \lbrack -1, 1] &\lbrack\;0.0, \;0.5] & $63.8\pm\;7.0$  \\
\enddata
\end{deluxetable}

\section{Discussion} \label{sec:discussion}
\subsection{Cross Check with Other Bulge Surveys} 

During 2015--2019, the OGLE-IV and MOA-II surveys conducted the bulge surveys, and since 2016 the KMTNet survey has been conducting its regular bulge survey. To double-check our microlensing event search, we cross-match our UKIRT events with the OGLE, MOA-II, and KMTNet events. Here we only check 2017--2019 events. 

\begin{figure}[ht!]
\includegraphics[width=0.48\textwidth]{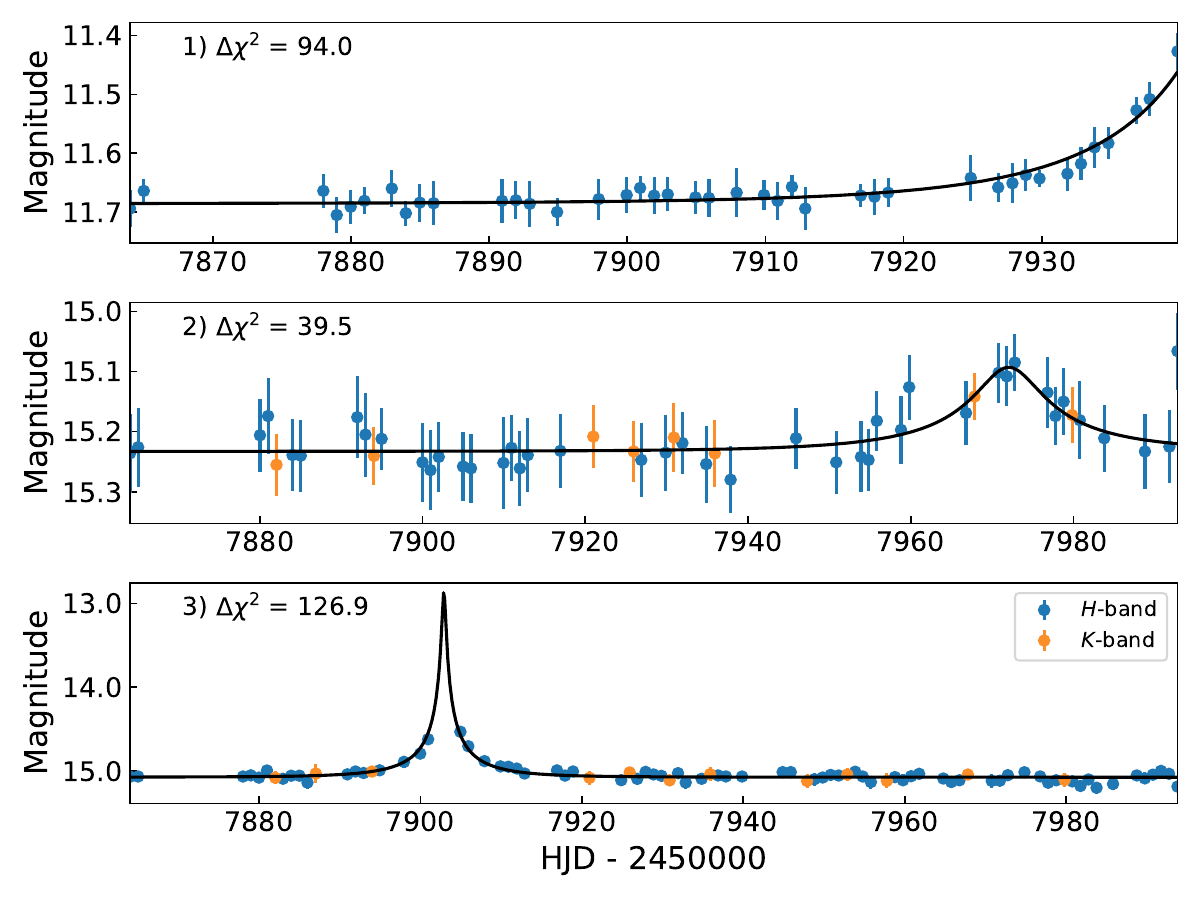}
\caption{Examples of KMTNet events missed by our search. These events were observed by both UKIRT and KMTNet but were missed by our UKIRT search due to the too small $\Delta\chi^2$. For the two events shown in the top panel and the bottom panel, the magnification peak occurred in the gap of the UKIRT observation window. For the event shown in the middle panel, the photometric accuracy and the cadence of the UKIRT survey are insufficient to detect it. The corresponding KMTNet event names are KB170714, KB170297, and KB170724 from top to bottom, respectively.}
\label{fig:kmtnet_miss_detection}
\end{figure}

During 2017--2019, there are a total of 8901 events reported by KMTNet \footnote{\url{https://kmtnet.kasi.re.kr/~ulens/}}, among which 442 events also have UKIRT observations. Our search identifies 67 of these events as clear events and 34 as possible events. The 341 remaining events are lost because their $\Delta\chi^2$ do not pass the $\Delta\chi^2 \geq 250$ criterion. Figure~\ref{fig:kmtnet_miss_detection} shows the light curves for three missing cases. We can confirm that all KMTNet events having UKIRT observations and satisfying our detection criteria have been detected by our search. 

We also conduct the same cross-matching analysis for events detected by OGLE and MOA.  During 2017--2019, there are a total of 5163 events reported by OGLE\footnote{\url{http://ogle.astrouw.edu.pl/ogle4/ews/ews.html}}, of which 394 also have the UKIRT observations. Our search labels 66 of these events as clear events and 30 as possible events. For MOA\footnote{\url{https://www.massey.ac.nz/~iabond/moa/alerts/}}, there are a total of 1376 events reported during 2017--2019, among which 96 events also have the UKIRT observations. Our search identifies 21 of these events as clear events and 11 as possible events. We can confirm that all OGLE and MOA events having UKIRT observations and satisfying our detection criteria have been detected by our search.

In addition, we find 11 possible anomalous events that were found by the OGLE, MOA, or KMTNet surveys. Table~\ref{tab:special_cases} summarizes the corresponding event names from the surveys. Because other surveys have higher cadences, a joint analysis of all survey data is encouraged. 

To facilitate future studies aimed at combing results from the UKIRT microlensing survey with those from other surveys, we have summarized events that have observations from the UKIRT survey and at least one other major microlensing survey in Table~\ref{tab:cross_match}. We also provide a data package containing these UKIRT light curves available online.

\begin{deluxetable*}{cccccc}
\tablewidth{0pt}
\caption{Summary of microlensing events with observations from UKIRT and at least one other microlensing survey (OGLE, MOA, or KMTNet).
The full table is available online. \label{tab:cross_match} }
\tablehead{
\colhead{sourceid} &  \colhead{RA} & \colhead{DEC} & \colhead{KMTNet ID} & \colhead{MOA ID} & \colhead{OGLE ID}  \\
\colhead{} & \colhead{(deg)} & \colhead{(deg)} &\colhead{} &\colhead{} &\colhead{}
}
\startdata
ukirt\_c\_2015\_n\_12\_3\_0075272&265.906832&-27.373336&-&-&OB151437\\
ukirt\_c\_2015\_n\_13\_1\_0052005&265.257974&-27.853824&-&-&OB150944\\
ukirt\_c\_2015\_n\_13\_4\_0063931&265.053514&-27.376525&-&-&OB151453\\
ukirt\_c\_2015\_n\_21\_2\_0008791&265.001610&-27.753371&-&-&OB151614\\
ukirt\_c\_2015\_n\_22\_2\_0064559&264.846871&-27.820285&-&-&OB151285\\
ukirt\_c\_2015\_n\_24\_4\_0013692&264.095165&-27.298911&-&-&OB151392\\
...&...&...&...&...&...
\enddata
\tablecomments{Events names of OGLE, MOA and KMTNet are abbreviations, e.g., OGLE-2015-BLG-1437 to OB151437.}
\end{deluxetable*}

\subsection{Future Work}

In this paper, we have searched for the microlensing events detected by the UKIRT microlensing survey. We have estimated the uncorrected microlensing event rates and found that there is probably a rising trend in the NIR event rates toward the Galactic center. The final goal of our work is to obtain the microlensing event rates in the NIR band toward the Galactic center, to help with the field selections for the NIR {\it Roman} microlensing survey. To achieve it, in the next paper (i.e., Paper II), we plan to conduct a detailed simulation of the detection efficiency for the UKIRT microlensing survey. Then, combining the events detected in this paper (i.e., Paper I), we can obtain the true NIR event rates toward the Galactic center. After that, we plan to apply the well-developed methodology to the VVV NIR survey and the ongoing NIR survey through the PRime-focus Infrared Microlensing Experiment \citep[PRIME;][]{Kondo23}. In addition, we may analyze the possible anomalous events to obtain the binary mass-ratio distributions in the direction toward the Galactic center.

\section{Conclusion} 
\label{sec:conclusion}

The 5-year UKIRT microlensing survey toward the Galactic center is designed for measuring the event rate and survey field selections for the {\it Roman} microlensing survey. Using the online UKIRT photometry data and the event-finder algorithm of \citet{Kim_2018}, we find a total of 522 clear events, 436 possible events, and 27 possible anomalous events. The model parameters of all PSPL events are derived using the MulensModel from \citep{Poleski_2019} and the emcee sampler code \citet{2013PASP..125..306F}. All the events have a median timescale $t_{\rm E}$ of $24.4$~days. Assuming a detection efficiency of $100\%$, we calculate the raw NIR event rate distribution near $|b|=0^{\circ}$, which should be the lower limit of the true NIR event rate. We find that there is probably a rising trend in the NIR event rates toward the Galactic center. All the data products from this work are made available online to the community. 

\section{Acknowledgments} 
We thank our anonymous referee for insightful comments, which help improve this manuscript a lot. We thank Yossi Shvartzvald for fruitful discussions. Y.W. and B.M. acknowledge the financial support from the National Key R$\&$D Program of China (2020YFC2201400), NSFC grant 12073092, 12103097, 12103098, the science research grants from the China Manned Space Project (No. CMS-CSST-2021-B09). W.Z. acknowledges the support from the Harvard-Smithsonian Center for Astrophysics through the CfA Fellowship.

This paper makes use of data from the UKIRT microlensing surveys (Shvartzvald et al. 2017) provided by the UKIRT Microlensing Team and services at the NASA Exoplanet Archive, which is operated by the California Institute of Technology, under contract with the National Aeronautics and Space Administration under the Exoplanet Exploration Program.

2015: UKIRT is currently owned by the University of Hawaii (UH) and operated by the UH Institute for Astronomy; operations are enabled through the cooperation of the East Asian Observatory. When the 2015 data reported here were acquired, UKIRT was supported by NASA and operated under an agreement among the University of Hawaii, the University of Arizona, and Lockheed Martin Advanced Technology Center; operations were enabled through the cooperation of the East Asian Observatory.

2016: UKIRT is currently owned by the University of Hawaii (UH) and operated by the UH Institute for Astronomy; operations are enabled through the cooperation of the East Asian Observatory. When the 2016 data reported here were acquired, UKIRT was supported by NASA and operated under an agreement among the University of Hawaii, the University of Arizona, and Lockheed Martin Advanced Technology Center; operations were enabled through the cooperation of the East Asian Observatory. We furthermore acknowledge the support from NASA HQ for the UKIRT observations in connection with K2C9.

2017: UKIRT is currently owned by the University of Hawaii (UH) and operated by the UH Institute for Astronomy; operations are enabled through the cooperation of the East Asian Observatory. When some of the 2017 data reported here were acquired, UKIRT was supported by NASA and operated under an agreement among the University of Hawaii, the University of Arizona, and Lockheed Martin Advanced Technology Center; operations were enabled through the cooperation of the East Asian Observatory. The collection of the 2017 data reported here was furthermore partially supported by NASA grants NNX17AD73G and NNG16PJ32C.

2018 - 2019: UKIRT is currently owned by the University of Hawaii (UH) and operated by the UH Institute for Astronomy; operations are enabled through the cooperation of the East Asian Observatory. The collection of the 2018 data reported here was supported by NASA grant NNG16PJ32C and JPL proposal 18-NUP2018-0016.

The authors wish to recognize and acknowledge the very significant cultural role and reverence that the summit of Mauna Kea has always had within the indigenous Hawaiian community. We are most fortunate to have the opportunity to use data produced from observations conducted on this mountain.

We thank the OGLE, MOA and KMTNet collaborations for making their event locations publicly available. 

Data DOI: 10.26133/NEA7




\bibliography{sample631}{}
\bibliographystyle{aasjournal}



\newpage
\appendix
\restartappendixnumbering

\section{Clear Events Detected}
\begin{figure*}[ht!]
\includegraphics[width=1\textwidth,height=1\textheight]{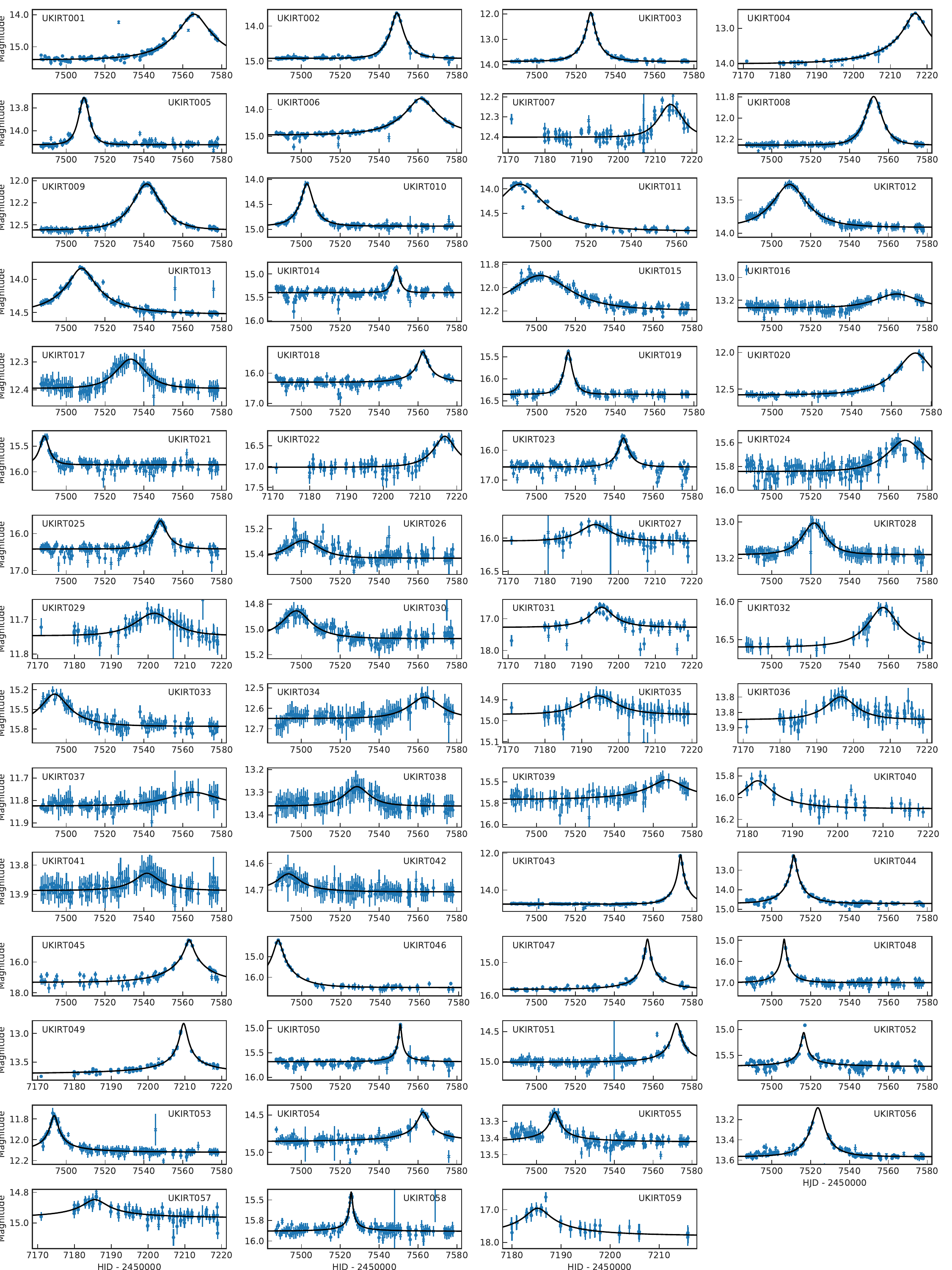}
\caption{Clear events detected from the 2015 to 2016 UKIRT microlensing survey data.
}

\label{fig:clear_events_2015-2016}
\end{figure*}

\begin{figure}
\figurenum{A2}
\plotone{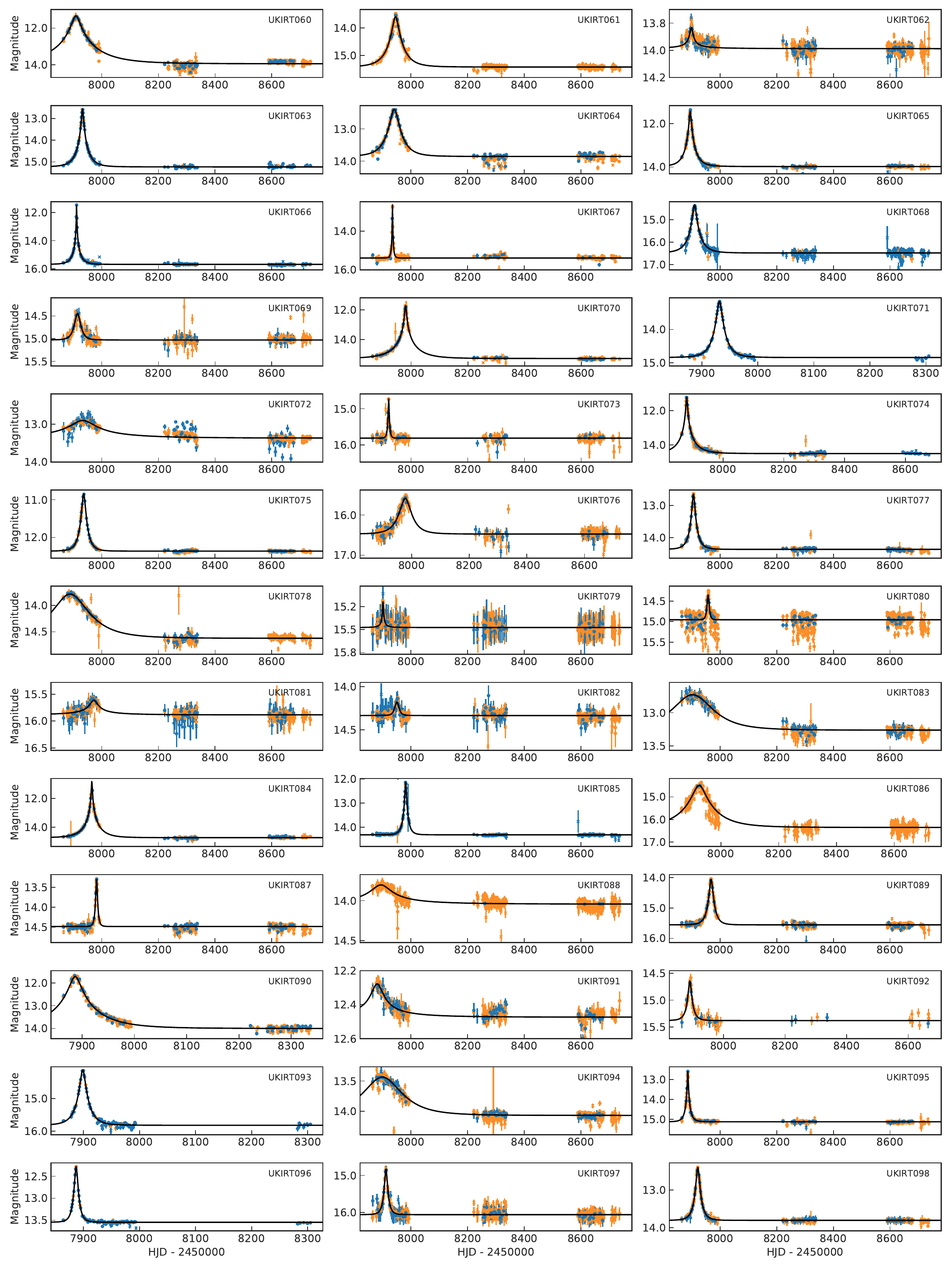}
\caption{Example clear events detected from the 2017 to 2019 UKIRT microlensing survey data. The H and K-band data are marked using blue and orange dots, respectively. The complete mosaic figure set is available in the online version of this Paper. \label{fig:clear_events_2017-2019.1}}

\end{figure}

In this appendix, we present light curves of all clear events detected from this work in Figure~\ref{fig:clear_events_2015-2016} to  ~\ref{fig:clear_events_2017-2019.1}. 
We have fitted a PSPL model to all these events light curves. The fitting parameters of all events are summarized in Table~\ref{tab:event_paramenter}.

\begin{deluxetable}{ccccccccc}
\tablewidth{0pt}
\caption{Parameters of the clear point source point lens microlensing events from the UKIRT survey, including coordinates and the best-fit parameters using a point source point lens model. Here we only show parameters for the first 5 events to save space.
\label{tab:event_paramenter} 
}
\tablehead{
\colhead{ID} & \colhead{RA} & \colhead{DEC} & \colhead{$u_0$} & \colhead{$t_0$} & \colhead{$t_{\rm E}$} & \colhead{$f_{bl}$}\\
& [deg] & [deg] & & (HJD-2450000)& (days) & 
}
\startdata
UKIRT001&269.331267&-28.070660&$0.270_{-0.009}^{+0.005}$&$7565.82_{-0.09}^{+0.09}$&$21.74_{-0.35}^{+0.55}$&$0.97_{-0.04}^{+0.02}$\\
UKIRT002&267.670632&-27.786195&$0.305_{-0.014}^{+0.008}$&$7549.15_{-0.03}^{+0.03}$&$7.80_{-0.12}^{+0.29}$&$0.96_{-0.06}^{+0.03}$\\
UKIRT003&268.146978&-28.058520&$0.148_{-0.010}^{+0.011}$&$7527.38_{-0.01}^{+0.01}$&$9.00_{-0.35}^{+0.37}$&$0.86_{-0.05}^{+0.06}$\\
UKIRT004&265.819739&-27.540654&$0.114_{-0.008}^{+0.009}$&$7216.72_{-0.03}^{+0.02}$&$19.81_{-1.11}^{+1.24}$&$0.35_{-0.03}^{+0.03}$\\
UKIRT005&269.670131&-28.000050&$0.691_{-0.072}^{+0.039}$&$7509.14_{-0.06}^{+0.06}$&$4.08_{-0.16}^{+0.29}$&$0.88_{-0.15}^{+0.09}$\\
...&...&...&...&...&...&...\\
\enddata
\end{deluxetable}

\figsetstart
\figsetnum{A2}
\figsettitle{Clear events detected from the 2017 to 2019 UKIRT microlensing survey data. }

\figsetgrpstart
\figsetgrpnum{A2.1}
\figsetgrptitle{Clear events detected from 2017 to 2019.}
\figsetplot{f13_1.pdf}
\figsetgrpnote{Clear events detected from the 2017 to 2019 UKIRT microlensing survey data.}
\figsetgrpend

\figsetgrpstart
\figsetgrpnum{A2.2}
\figsetgrptitle{Clear events detected from 2017 to 2019.}
\figsetplot{f13_2.pdf}
\figsetgrpnote{Clear events detected from the 2017 to 2019 UKIRT microlensing survey data.}
\figsetgrpend

\figsetgrpstart
\figsetgrpnum{A2.3}
\figsetgrptitle{Clear events detected from 2017 to 2019.}
\figsetplot{f13_3.pdf}
\figsetgrpnote{Clear events detected from the 2017 to 2019 UKIRT microlensing survey data.}
\figsetgrpend

\figsetgrpstart
\figsetgrpnum{A2.4}
\figsetgrptitle{Clear events detected from 2017 to 2019.}
\figsetplot{f13_4.pdf}
\figsetgrpnote{Clear events detected from the 2017 to 2019 UKIRT microlensing survey data.}
\figsetgrpend

\figsetgrpstart
\figsetgrpnum{A2.5}
\figsetgrptitle{Clear events detected from 2017 to 2019.}
\figsetplot{f13_5.pdf}
\figsetgrpnote{Clear events detected from the 2017 to 2019 UKIRT microlensing survey data.}
\figsetgrpend

\figsetgrpstart
\figsetgrpnum{A2.6}
\figsetgrptitle{Clear events detected from 2017 to 2019.}
\figsetplot{f13_6.pdf}
\figsetgrpnote{Clear events detected from the 2017 to 2019 UKIRT microlensing survey data.}
\figsetgrpend

\figsetgrpstart
\figsetgrpnum{A2.7}
\figsetgrptitle{Clear events detected from 2017 to 2019.}
\figsetplot{f13_7.pdf}
\figsetgrpnote{Clear events detected from the 2017 to 2019 UKIRT microlensing survey data.}
\figsetgrpend

\figsetgrpstart
\figsetgrpnum{A2.8}
\figsetgrptitle{Clear events detected from 2017 to 2019.}
\figsetplot{f13_8.pdf}
\figsetgrpnote{Clear events detected from the 2017 to 2019 UKIRT microlensing survey data.}
\figsetgrpend

\figsetgrpstart
\figsetgrpnum{A2.9}
\figsetgrptitle{Clear events detected from 2017 to 2019.}
\figsetplot{f13_9.pdf}
\figsetgrpnote{Clear events detected from the 2017 to 2019 UKIRT microlensing survey data.}
\figsetgrpend

\figsetgrpstart
\figsetgrpnum{A2.10}
\figsetgrptitle{Clear events detected from 2017 to 2019.}
\figsetplot{f13_10.pdf}
\figsetgrpnote{Clear events detected from the 2017 to 2019 UKIRT microlensing survey data.}
\figsetgrpend

\figsetgrpstart
\figsetgrpnum{A2.11}
\figsetgrptitle{Clear events detected from 2017 to 2019.}
\figsetplot{f13_11.pdf}
\figsetgrpnote{Clear events detected from the 2017 to 2019 UKIRT microlensing survey data.}
\figsetgrpend

\figsetgrpstart
\figsetgrpnum{A2.12}
\figsetgrptitle{Clear events detected from 2017 to 2019.}
\figsetplot{f13_12.pdf}
\figsetgrpnote{Clear events detected from the 2017 to 2019 UKIRT microlensing survey data.}
\figsetgrpend

\figsetgrpstart
\figsetgrpnum{A2.13}
\figsetgrptitle{Clear events detected from 2017 to 2019.}
\figsetplot{f13_13.pdf}
\figsetgrpnote{Clear events detected from the 2017 to 2019 UKIRT microlensing survey data.}
\figsetgrpend

\figsetend

\end{document}